# Comparative Assessment of Markov Models and Recurrent Neural Networks for Jazz Music Generation

**Conrad Hsu[1], Ross Greer[2]**
[1] Carlmont High School, Belmont, CA
[2] Electrical & Computer Engineering, University of California San Diego, La Jolla, CA

**Student Authors**
Conrad Hsu (High School)

**SUMMARY**

As generative models have risen in popularity, a domain that has risen alongside is generative models for music. Our study aims to compare the performance of a simple Markov chain model and a recurrent neural network (RNN) model, two popular models for sequence generating tasks, in jazz music improvisation. While music, especially jazz, remains subjective in telling whether a composition is "good" or "bad", we aim to quantify our results using metrics of groove pattern similarity and pitch class histogram entropy. We trained both models using transcriptions of jazz blues choruses from professional jazz players, and also fed musical jazz seeds to help give our model some context in beginning the generation. Our results show that the RNN outperforms the Markov model on both of our metrics, indicating better rhythmic consistency and tonal stability in the generated music. Through the use of music21 library, we tokenized our jazz dataset into pitches and durations that our model could interpret and train on. Our findings contribute to the growing field of AI-generated music, highlighting the important use of metrics to assess generation quality. Future work includes expanding the dataset of MIDI files to a larger scale, conducting human surveys for subjective evaluations, and incorporating additional metrics to address the challenge of subjectivity in music evaluation. Our study provides valuable insight into the use of recurrent neural networks for sequential based tasks like generating music.

**INTRODUCTION**

In recent years, the popularity of generative models has surged in parallel with the widespread adoption and growing fascination surrounding Artificial Intelligence (AI). Particularly with the recent development of ChatGPT and DALL- E, researchers are even more fascinated by the ability of these models to generate an answer to almost anything asked [1]. However, with generative models being applied to text such as through Natural Language Processing (NLP) or image classification and generation, a field that remains relatively under-examined compared to the rest of its counterparts is music generation.

Developing generative models for music is a complex undertaking due to the inherent subjectivity of the field and the multitude of approaches available for training such models. Evaluating the quality of generated music poses a significant challenge, given its subjective nature. Nonetheless, music encompasses a theoretical foundation comprising elements like scales, keys, and chords, which can be effectively identified by machine learning models. However, the elusive aspects of consciousness and soul, integral to human musicians, present a persistent hurdle in achieving truly captivating generative music. While AI algorithms can analyze vast amounts of musical patterns and data [2], they struggle to capture the intricate subtleties and emotions that only a human brain is capable of. This depth of emotional connection and the ability for AI models to create nuanced musical ideas remains a difficult challenge.

When thinking about music, melodies can be framed as a sequence or series of events. In this sequence, different musical notes transition to other musical notes, each note being sustained for a certain duration, and being played with varying degrees of intensity or velocity. These musical notes, which are the building blocks for music, have different characteristics such as pitch and duration. With this in mind, a simple Markov chain is a good place to start when trying to generate music, as a Markov chain is essentially a set of transitions determined by some probability distribution [3]. In addition, another go-to neural architecture for generating music is the Recurrent Neural Network (RNN). Its popularity stems from being a deep learning algorithm that is specialized for processing sequential data [3]. In fact, a main advantage is its ability to share weights across time, meaning it can remember previous inputs and make more accurate outputs compared to a standard neural network.

However, these models are unable to process data straight from audio or a piece of sheet music. Tokenization, which is the start of the NLP process, has been frequently used in text classification, virtual chatbots, sentiment analysis, and many other applications.

Tokenization can be thought of as breaking raw texts into smaller chunks called tokens [4]. In the context of generative models for music, tokenization allows for a finer level of granularity to analyze and generate the music. In our research, we tokenize notes from different inputted Musical Instrument Digital Interface (MIDI) files in order to break each specific attribute of a note such as pitch and duration.

Our project aims to create both a Markov chain model and a RNN model to generate music from the jazz genre. We hypothesize that a recurrent neural network model outperforms a simple Markov model in improvising jazz music on the metrics of grooving pattern similarity [4] and pitch class histogram entropy [5]. While jazz music has its theoretical rules similar to the rules of harmony and counterpoint that guide classical composition, jazz has a more free-flowing element with the idea of improvisation. Within this genre of music, we can test the generative capabilities of a traditional Markov model compared to an RNN model. In this study, we evaluate our models through a series of metrics called pitch class histogram entropy and grooving pattern similarity, which both aim to analyze and quantify certain expected similarities between human-generated and AI-generated music. The pitch class histogram measures the distribution of pitches within a piece of music, while the grooving patterns similarity helps to assess the continuity and rhythmicity of the music. In our study, we found that the RNN did outperform the Markov model on both metrics and discussed the further implications of solely using these metrics to evaluate the quality of the generated jazz improvisations. We make our code publicly available at https://github.com/Chsu123/Jazz-Improv-Models

**RESULTS**

We investigated how well two different models, a Markov model and a RNN model, could improvise jazz. We trained both our models on the same set of 20 MIDI files, each MIDI file being a monophonic, single line, 12 bar blue transcribed from professional jazz players. Using music21 [6], a Python library for computational music analysis, we tokenized or converted each musical note and rest in the MIDI files into tokens that our models are able to interpret.

For our Markov model, the model takes all 20 MIDI files and iterates over each one to extract the unique pitches and durations, helping the model capture the unique musical patterns in our dataset. To build the transition table, the occurrences of musical notes following each sequence of previous notes are counted. The transition table is structured as a nested dictionary, where each sequence of previous notes (current state) maps to a dictionary of next notes and their occurrence counts. To calculate transition probabilities, the occurrence counts of the next notes are divided by the total count of all possible next notes for each current state, resulting in normalized probabilities. These probabilities represent the likelihood of transitioning from one note sequence to another, providing a basis for generating new music **(Figure 1)**.

In order to help "jump start" the model or instead of letting our model choose a random pitch and duration to start off on, we utilized a unique 16-note jazz seed as the initial input for each generated output. This approach ensured that the models would have the best chance at generating an authentic jazz improvisation, as both our Markov and RNN model received the same set of 16-note jazz seed pairs. By using these predetermined seeds, it also helped to establish a more common reference for comparison between the two models.

For our RNN model, the model also took all 20 MIDI files and iterated over them to extract sequences of notes from MIDI files, tokenizing them so the model can interpret these sequences. Our neural network is then trained on these sequences for 200 epochs and saves the best weights based on the loss value during training. We then pass the same 16 note seeds we used for the Markov model into the RNN, and our model then generates a sequence of notes based on the given seed sequence. Similar to the Markov model, it selects the most probable next note at each step during the model's predictions. Our model then took the newly generated composition and converted the sequences of notes into a MIDI file, which we could download and analyze.

While we could simply listen to the generated outputs from each model as a means of subjective evaluation, we aimed to compare the models using a more metrified way such as the groove pattern similarity metric and the pitch class histogram entropy metric. Groove pattern similarity quantifies the rhythmic quality and overall swing field of jazz music by capturing the positions in the bar where notes onsets (where musical notes are played) and comparing those onsets to other onsets within the composition. On the contrary, the pitch class histogram, which doesn't focus on the rhythmic integrity, simply measures the distribution of pitches within the composition and calculates the overall entropy of the pitch distribution.

Our RNN outperformed our Markov model on both metrics, with the RNN having a higher groove pattern metric 62.5 percent of the time compared to the Markov model **(Figure 2)**. The RNN also outperformed the Markov model 100 percent of the time on the pitch class histogram entropy test **(Figure 2)**, with a lower entropy marking a clearer tonality, which ideally meant a more melodic generated composition.

In our 6th seed composition, the RNN outperformed the Markov model in terms of pitch distribution. The pitch stability was higher, with the composition being centered around less notes compared to the Markov model, which was centered around more pitches, thus increasing the entropy **(Figure 3)**. Furthermore, looking at the composition seed 1, this was where the Markov model had outperformed the RNN on the groove pattern similarity metric **(Figure 2)**. The Markov model showed more rhythmic stability with a steadier trend line compared to the RNN model **(Figure 4)**.

Looking at a MIDI representation of our outputs for the seed 1, the Markov model clearly shows more simplicity in rhythms **(Figure 5a)**, hence its better groove pattern similarity. The MIDI representation for the RNN shows a greater rhythmic complexity, with more variety after the 16 note seed sequence **(Figure 5b),** hence a higher entropy score for the grooving metric. However, despite the higher variety of rhythms, the pitches of the notes stay rather near each other **(Figure 5b)**, instead of the Markov model, which make greater pitch jumps between quarter notes and half notes **(Figure 5a)**, hence the RNN being superior in the pitch class histogram entropy metric.

**DISCUSSION**

Our experiments support our hypothesis that RNNs would outperform the Markov model for jazz composition generation on the metrics of grooving pattern similarity and the pitch class histogram metric. These findings match the current trends in the direction of the music and AI field, as neural networks appear to create better music than simple generation schemes like the Markov model [7].

The Markov model's architecture, which involves simply calculating the highest probability based on the durations and pitch of the notes at a specific moment (**Figure 1**), results in its lack of long term dependency. This means that while the melodies might be simpler as every note and duration is mapped to a specific chance of transitioning to another note, this results in non-melodic transitions at unexpected moments. This is clearly shown in the results of the pitch class histogram entropy metric (**Figure 2**), as the Markov model tended to deviate from the general pitches. In the context of music theory, for any genre of music, the best sounding melodies tend to have transitions between notes that make sense, rather than jumping from one octave to the other in the middle of a potentially melodic phrase. The MIDI file of the model generations (**Figure 5)** demonstrates this as after the planted seed, in which the first 16 notes of the generation are exactly the same, the Markov model takes a simpler approach by just including half notes and quarter notes. While it doesn't make huge pitch jumps as the model is well aware of the most common probabilities trained from professional jazz recordings, the choice of notes jump to different notes too often, with coming back to them. In jazz theory, the idea of enclosing a note, which means hanging around a certain target note, leads to a more authentic jazz sound. The RNN (**Figure 5b)** embodies the idea of enclosing notes as seen in the note pitches going up and down, enclosing notes back and forth, which helps to keep the output more authentic by keeping the pitch more stable. The RNN is able to capture enclosures better than the Markov model due to the long short term memory (LSTM) architecture of the neural network. The hierarchical structure of LSTM helps to capture the essence of jazz by learning patterns at different levels of abstraction. The lower-level LSTM layers capture-short term patterns within the training MIDIs which might include small melodic motifs, such as the ideas of enclosures within these mini-motifs the models learn. On the other hand, the higher-level LSTM layers capture the long term patterns that help the smaller melodic motifs learn to remain in context, or not be generated at random spots that don't make musical sense. Thus, the pitch class histogram entropy metric (**Figure 2**) was accurate in helping to show the RNN outperforming the Markov model in terms of a lower pitch entropy or more stability, which led to more authentic jazz being generated.

      While the RNN always offered greater pitch stability compared to the Markov model, there were discrepancies between the results when it came to the groove pattern similarity test (**Figure 2)**. This could partly be explained by the idea that the groove pattern similarity metric was designed to evaluate the rhythmic consistency within the context of the generated piece and not in terms of the rhythmic consistency of jazz. Thus, due to the simplicity of the Markov model's architecture, the Markov model, which is infamous for its ability to get stuck in a loop, performs relatively well on the groove pattern metric, which simply measures the onsets of notes within every consecutive measure. As seen in the seed 1 pair of generations, this is an instance when the Markov model outperforms the RNN on the groove pattern similarity test (**Figure 2)**. This can be seen as a result of the note onsets being more similar to each other, as the notes where the beat lands between the consecutive measures is more similar (**Figure 5a)**, compared to the RNN output (**Figure 5b)**, which shows less similarity in where the note onsets occur, as there involves many more complicated rhythmic patterns, thus, less notes land on the same beat within each consecutive measure. This explains the idea of the RNN not outperforming the Markov model every time on this metric, as the Markov model actually has an advantage and greater chance with the note onsets between each measure being more similar, as there is less variation in the output of the Markov model, as it doesn't capture intricacies as well as an RNN. Yet, the architecture of the RNN still helped the RNN to beat the Markov model

on this metric. The use of activation functions within our RNN such as (Rectified Linear Unit) helped to introduce non-linearities that allowed for more complex mappings between the input and output. The dense layers helped to capture high-level learning representations of the training data, which helped generate compositions more similar to those of professional jazz musicians, who are able to create melodic melodies due to their ability to keep the note onsets between each measure similar to each other. Thus, the RNN is able to create intricate melodies with the use of layers, while putting these melodies in context, through the use of making the onesets as similar as possible within each consecutive measure, explaining the greater groove pattern similarity metric score.

While the Markov model performed on a similar level to the RNN in terms of groove pattern similarity, this could be explained by the idea that Markov models lack the ability to create more complicated and interesting melodies, as seen by the MIDI file (**Figure 5).** The purpose of the groove pattern metric is to quantify the idea of "grooving" in jazz, which refers to the quality of persisted rhythms that give a sense of legitimacy. The famous jazz mantra "repetition legitimizes" greatly supports this idea. Therefore, when personally listening to the audio generations, we felt that the Markov model also did a great job encapsulating the authenticity of jazz, as the simpler melodies and rhythms did help add to the idea of a "groove feel". Nonetheless, the subjectivity of music makes classifying music as "good" or "bad" extremely difficult. When listening to the RNNs, while sometimes the compositions felt more chaotic with more variety in the rhythms, but the compositions were far more interesting from a jazz theory perspective – it encapsulated musical ideas such as modulation and rhythmic complexity. The use of dropout between the dense layers in our RNN helped to prevent overfitting, thus reducing the model's reliance on specific patterns which helps to improve the overall generalization. However, we thought the Markov model was better at producing more melodic melodies, as more often than not, simple melodies are catchier, thus making it more appealing.

In addition, a limitation of the groove pattern metric was that it captured note onsets within consecutive measures. This meant that a composition that had only quarter notes for every beat would earn a perfect score of 1 on this metric; however, this clearly wouldn't make a good jazz composition and would certainly not be "groovy". The groove pattern metric provides a quantitative measure to assess the overall similarity of rhythms, but it shouldn't be used solely to judge how well a composition grooves. Similarly, the limitation of the pitch class histogram entropy metric stems from the idea that if the same pitch is played throughout the whole piece, the entropy calculation would result in a score of 0 due to the no changes in variation. As a result, a composition of playing one note at the same pitch and duration would earn perfect scores on our jazz metrics. This concept adds on to why it is so difficult to evaluate how well a model generates music, other than using humans as a source. Thus, we hope to improve this study in the future by conducting human survey responses to determine which model is perceived as producing better jazz music. Nonetheless, these metrics provide a quantitative measure of how well our models performed, as the outputs from both models were random enough. Thus, apart from aural assessment, the utilization of these straightforward metrics was effective in evaluating pitch and rhythmic integrity, as these two things usually mark a better sounding piece, in any genre of music.

Other future experiments may include training both models on a larger dataset, as a limitation of our research was only training both on 20 MIDI files of a blues. While our models were still able to output music that was jazz-like, a larger dataset could have resulted in more authentic sounding jazz. Other metrics we hope to use to help compare the results of our Markov and RNN model include the MIREX-like continuation challenge [4], which involve seeing how well a model can continue a melody when given tiny excerpts from the training data. This could help to provide a more accurate approach at assessing how well these models hold up against the professional recordings of jazz.

**MATERIALS AND METHODS**

First, we generated our dataset of 20 jazz samples from blues by manually transcribing a choruses (1 blues chorus = 12 bars) from professional players Charlie Parker, Emmet Cohen, Kenny Barron, Oscar Peterson, and Wes Montgomery. We transcribed the monophonic lines of their solos and kept the transcriptions in MuseScore, a music notation software.

We then created our model architectures, with all coding being done in a Google Colab notebook. To build the Markov model, we utilized music21's library to help convert the 20 jazz MIDI files into tokens, of pitch and duration (e.g."C4_0.5" represents a C4 quarter note). Python Collections was also utilized to help us work with ordered dictionaries in storing our notes in the sequential order they were found in from the training dataset. To build our transition probability table, we chose a sequence number of 3, meaning the model examines a group of 3 consecutive notes as the current state and records the subsequent note that follows this state (in other words, an order-3 Markov model). This occurrence of the next state following the current state is recorded in a transition table, and the probabilities are normalized. The total count of all possible next states for each current state is calculated, and then the counts are divided by the total count to obtain probabilities. To create the 8 pairs of musical seeds we used to help give our models a more equal comparison, we customly created 16 note jazz improvisation starters to feed into our models. After passing a seed file path to our model, it is tokenized and stored as a list to be used as the initial state. The Markov model then generates new music by using the seed as the initial state and looks at the current state and consults the transition table to select the note based on the highest probabilities. This process was repeated until 200 new notes were generated, which was a number that gave us a decent chunk of music(approximately 30 measures) to look at.

      To create our RNN model, Tensorflow, music21, Keras, Collections, and Numpy libraries were imported. The network architecture was implemented using Keras with LSTM layers, batch normalization, and dropout to prevent overfitting. The network had two LSTM layers, each with 512 units, and two Dense layers with 256 and n_vocab (representing the number of unique notes and chords) units, respectively. The model was trained using the Adam optimizer with a categorical cross-entropy loss function. A model checkpoint callback was used from Keras to save the best weights with the minimum loss, and these weights were updated using backpropagation. The temperature parameter was introduced to control the randomness of the network's predictions during music generation. A value of 1 for temperature ensured a moderate level of randomness, while lower values led to more deterministic predictions. The model was trained for 200 epochs with a batch size of 64 and the generated music sequences were saved. We then created a function to extract the 16 note seed from the provided MIDI file, with the pitches and duration being converted into an integer index and one-hot encoding is applied to transform the integer indices into binary arrays that represent the musical elements. The function responsible for generating the musical output operates on the RNN model, network input (encoded seed sequence), and vocabulary of possible musical elements. It iteratively predicts the next musical element using the RNN model based on the seed sequence, generating an extended musical sequence over 250 steps. To introduce randomness and diversify the output, the numpy.random.choice method is utilized during the prediction process. After the prediction output is saved, each element of pitch and duration is extracted and music21 saves the prediction output in the final output. Stream from music21 is then called to convert the final output into a downloadable MIDI file that was saved into our notebook directory.

For our jazz metrics, groove pattern similarity denoted as $GS(\vec{g_a}, \vec{g_b})$, is a mathematical measure used to quantify the rhythmicity of music. The grooving pattern $\vec{g}$ represents a 64-dimensional binary vector, where each dimension corresponds to a position in a bar, and a value of 1 indicates the presence of a note onset at that position. The similarity between two grooving patterns, $\vec{g_a}$ and $\vec{g_b}$, is calculated by applying the exclusive OR (XOR) operation on their corresponding dimensions, and then taking the average of the resulting vector. The formula is denoted as:

$$GS(\vec{g_a}, \vec{g_b}) = 1 - \frac{1}{Q} \sum_{i=0}^{Q-1} XOR(g_i^a, g_i^b)$$

where G is the dimensionality of the grooving patterns, and $g_i^a$ and $g_i^b$ represent the i-th dimensions of the patterns. Our code for the groove pattern similarity metric loads the generated MIDI files and iterates over the bars, extracting the musical elements such as notes and rests. For each bar, a 64-dimensional binary grooving pattern represents the note onsets within that bar. Subsequently, the similarity between all consecutive bars is calculated which applies the $XOR$ operation on their respective grooving patterns and then calculates the average similarity score after calculating the score between all pairs of adjacent bars in the composition. We then imported Matplotlib to create a line graph (**Figure 4)** that plotted the relationship between the bar pairs and similarity scores.

For our pitch class histogram entropy metric, which analyzed the different pitches, the code utilizes a 12-dimensional pitch class histogram, represented by the vector $\vec{h}$, to collect the notes appearing within a specific period, such as a bar. The histogram is constructed based on the 12 pitch classes (C, C#, D, D#, E, F, F#, G, G#, A, A#, B), and the frequency of each pitch class is normalized by dividing it by the total note count in the period, resulting in the sum of the probability distribution being equal to 1. The entropy of $\vec{h}$ is then calculated using the formula:

$$H(\vec{h}) = -\sum_{0}^{11} h_i log_2(h_i)$$

where the sum is taken from all 12 pitches. Our code simply extracts all notes belonging to the corresponding pitch class from the inputted MIDI file. We then created a bar plot from Matplotlib (**Figure 3)** to plot the frequency of each pitch with the corresponding pitch class.

**FIGURES, TABLES, AND CAPTIONS**

```
Current state: ('D5_1.0', 'C5_0.5', 'D5_0.5')
Next state: C5_0.5 Probability: 1.0
-------------------
Current state: ('C5_0.5', 'D5_0.5', 'C5_0.5')
Next state: D5_1.0 Probability: 1.0
-------------------
Current state: ('D5_0.5', 'C5_0.5', 'D5_1.0')
Next state: C5_0.5 Probability: 1.0
-------------------
Current state: ('C5_0.5', 'D5_1.0', 'C5_0.5')
Next state: D5_1.0 Probability: 1.0
-------------------
Current state: ('D5_1.0', 'C5_0.5', 'D5_1.0')
Next state: F5_0.5 Probability: 1.0
-------------------
Current state: ('C5_0.5', 'D5_1.0', 'F5_0.5')
Next state: D5_0.5 Probability: 1.0
-------------------
Current state: ('D5_1.0', 'F5_0.5', 'D5_0.5')
Next state: D5_1/6 Probability: 1.0
-------------------
Current state: ('F5_0.5', 'D5_0.5', 'D5_1/6')
Next state: A4_2/3 Probability: 1.0
-------------------
Current state: ('D5_0.5', 'D5_1/6', 'A4_2/3')
Next state: F4_0.5 Probability: 1.0
-------------------
```

**Figure 1**. **Markov Probability Transition Table.** Sample of generated transition probabilities that guide model in predicting the next note, based on the current state of 3 notes. The generated output of Markov model is generated from these probabilities; we note that in this particular example, due to the 3rd-order Markov model containing relatively sparse instances of these particular 3-note sequences, the probability of the next note is always 1. Other sequences that appear more frequently in the training data (with more diverse next-note destinations) will have more variance in the next-note probabilities.

|  | **Markov Model** | | **RNN** | |
| --- | --- | --- | --- | --- |
| **Composition Seed** | **Groove Pattern Similarity** | **Pitch Class Histogram Entropy** | **Groove Pattern Similarity** | **Pitch Class Histogram Entropy** |
| 1 | **0.947** | 3.429 | 0.926 | **3.199** |
| 2 | 0.939 | 3.379 | **0.951** | **3.152** |
| 3 | 0.947 | 3.391 | **0.954** | **3.182** |
| 4 | 0.945 | 3.426 | **0.955** | **3.125** |
| 5 | 0.949 | 3.291 | 0.906 | **3.140** |
| 6 | 0.9363 | 3.317 | **0.944** | **3.074** |
| 7 | **0.959** | 3.324 | 0.944 | **3.086** |
| 8 | 0.958 | 3.327 | **0.969** | **3.216** |

**Figure 2. Table of Groove Pattern Similarity and Pitch Class Histogram Entropy Scores for Markov and RNN Models.** Groove pattern similarity scores range from 0 to 1, where a score closer to 1 indicates a higher level of similarity between all adjacent bars in terms of rhythmic onset and 0 being completely dissimilar. Higher pitch class histogram entropy score indicates greater pitch diversity in composition and lower score indicating a more focused tonal center.

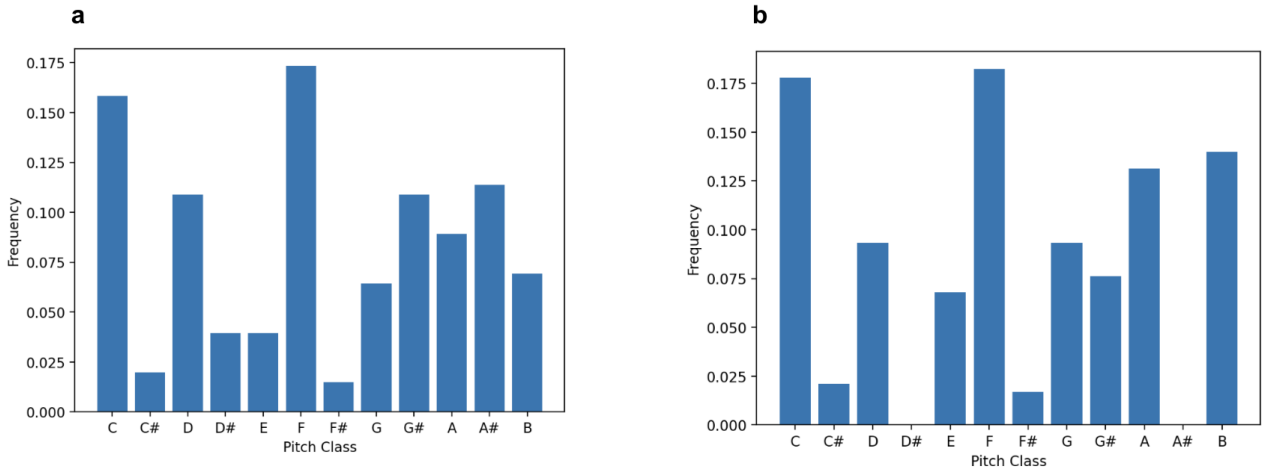

**Figure 3. Barplot of Pitch Class Histogram for: a) Seed 6 Markov Composition and b) Seed 6 RNN Composition**. Barplot shows each of the 12 pitches (C, C#, D, etc…) and the frequency percentages of how often the pitch appears in the composition, with total frequencies adding up to 1.

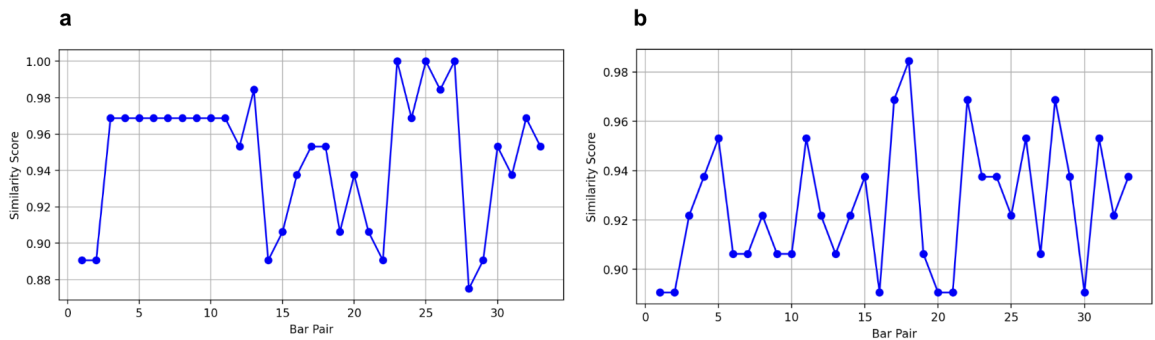

**Figure 4. Line Graph of Groove Pattern Similarity Between all Adjacent Bars for: a) Seed 1 Markov Composition and b) Seed 1 RNN Composition.** Line plot illustrates similarity

scores for each pair of adjacent bars. On the x–axis, 1 indicates the first bar pair between measure 1 and 2, 2 indicates second bar pair between measure 2 and 3, etc…

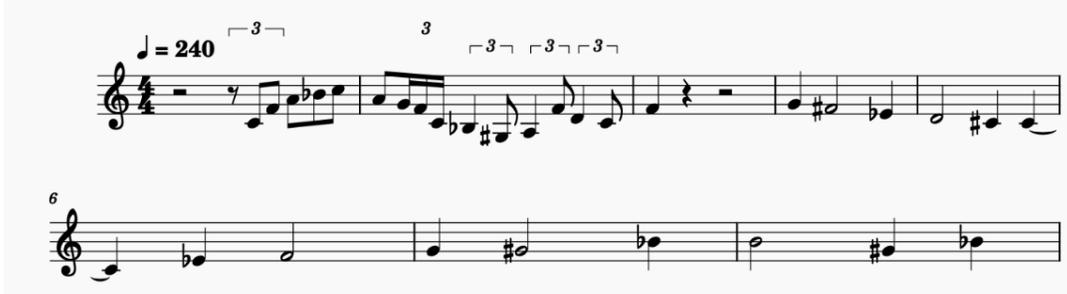

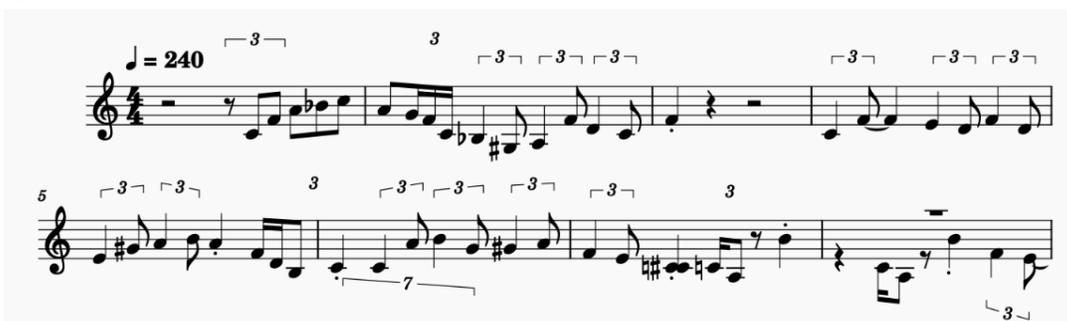

**Figure 5. MIDI representation for: a) Seed 1 Markov Composition and b) Seed 1 RNN Composition.** First 8 measures of Seed 1 composition for both models, showcasing the musical notation of MIDI, which was notated in MuseScore.